Packaging "Spectra" (as in Partition Functions and L/ζ-functions) to Reveal Symmetries (Reciprocity) in Nature and in Numbers

Martin H. Krieger krieger@usc.edu

NOTE <u>Symmetry</u> means that some feature does not matter, so circular symmetry means that the compass direction does not matter. If that symmetry is broken or violated, then the system is now <u>ordered</u>, in this case it points in a particular direction. Note that if it points in a particular direction, it may be said to be symmetric if the system is rotated around that compass direction (but note that the rotation cuts through the plane of the compass.

Geometry and Harmony

The world as we see it and experience it is <u>geometrical</u>. Maxwell in about 1860, earlier actually ??Faraday, understood this to say that to understand geometry is to understand the forces of nature. Here we might think of a magnet and its lines of force revealed by iron filings.

There is a corresponding <u>harmony</u> of nature, so that shape determines sound, here think of a musical instrument, the shape and size of a drum determining its sound spectrum. Conversely, those harmonies when considered as a whole, point to their <u>source</u>, and any orderliness we find in that harmony tells about the deep features of that source. So atomic spectra, revealed when we heat up matter to a gas, tell us about the structure of the atoms that compose that matter.

We might think of that harmony as a <u>package</u> of information about the source, and often features of that package (say symmetries) tell us otherwise less apparent features of the source.

We may discern that other "spectra," such as the prime numbers pointing to their "source" and its structure, and elementary particles (and their properties) pointing to the fundamental forces of nature and their structure. Or, considering the sound of a drum, we might package those sounds into a function that will tell us about its shape (Weyl asymptotics, Kac on hearing the shape of a drum); or we might think of the Greens function of quantum electrodynamics, as defined by Schwinger, as packaging the various Feynman paths and diagrams whose sum leads to the amplitude for a process. And rather more generally, the Selberg trace formula.

Of course, we'll need to understand what it means to be an "of course," and how we are "pointing to" it. Those structures have symmetries, shapes so to speak, not at all manifest from the individual objects, the primes, particles, or tones—those symmetries are in effect emergent from considering the harmony as a whole.

Parts and the Right Parts

A bit of matter is composed of atoms and molecules, much as a number is a product of the prime numbers (12=2x2x3). It is essential that we have the <u>right parts</u>, so the composition makes sense and is useful (to think with?). So we might decompose something into its more elementary parts (and those parts may then be decomposed). We'll discern the rules of composition, whether they be additive, one-by-one, or multiplicative, or the more complex rules of musical composition. Or, they may be parts that might be seen as <u>components</u> of the whole, where those parts may be parts of an automobile or organs of animals. Namely, there are parts that compose a more complicated whole, those parts are themselves composed of even more elementary parts.  Each part on its own scale has the integrity of an individual, a carburetor, an ingredient in a recipe. In general, larger more encompassing parts are more specific than their components. They are in effect more ordered, but they are more symmetrical than the things they compose.

Plenitude

Given a set of parts (numbers, atoms, people) there are rules for how they are to be put together, but given those rules <u>any combination of parts</u> is allowed and surely is instantiated. Hence, the variety of recipes in a cookbook, the chemicals we see in nature, the organization of a society in terms of its marriage and kinship rules or in terms of its economy (what is exchangeable for what).

By the way, we might not explicitly appreciate those rules of composition.
We discover more about those rules when new compositions prove successful or are drastically wrong. We may then discern what the rules allow, and what they forbid. So for example, many notional schemes of power production violate the conservation of energy, and they are shown to be impractical.

Manifold Profiles or Perspectives

There is value in having <u>diverse ways of thinking</u> about nature, even if in the end they come to the same result. One way may build in orderliness or symmetry in an explicit way. Another may attend in detail to the components of what you are studying. So that same result may be shown to have orderliness built in, or it may be show how the different parts work together to get that same result.

For example, we can package the prime numbers, $p_i$, into a function $\zeta(s) = \Pi (1-p_i^{-s})^{-1}$ (where $\Pi$ means product). We can show that that function (of $s$), packaging the primes, is related (through a Laplace-like Mellin transform) to another function $\theta$, which plays a role in understanding the flow of heat, where the primes would seem to play no role. (And the transform would seem not to introduce the primes behind our backs.) Now $\theta$ has all sorts of lovely symmetries that make sense when we think of heat flow. (And in thinking of heat flow, we realize that it takes place by molecules colliding with each other, what is called a random walk, as a model of diffusion of heat, and that introduces new symmetries into $\theta$.)

Another example: imagine a two-dimensional grid with microscopic magnets at each vertex. One way of computing the function that tells us how the grid behaves takes into account each of those magnets and discovers the right objects that may be added up to understand the behavior of the grid. (The right parts here are not the individual magnets.) Another way, explores the various symmetries of the grid, for example a pair of magnets that are aligned when the grid is at a high temperature, is analogized to a pair of magnets that are anti-aligned at a very low temperature. Both ways of computing  come to the same result, as we might hope. So we might discover that the first way builds in those symmetries, although they are were not explicitly included.

We have manifold perspectives on an identity, each perspective telling us a distinctive feature of that identity that was not manifest from the other perspectives. Here counting or adding-up is complemented by symmetry of the whole object.

Again, features of the world that might be difficult to discern from one perspective are manifest from another. If may be difficult to tell if a number is prime, especially if it is large, but from another perspective there is a straightforward and efficient procedure for doing so. In the theory of numbers in mathematics, such a relationship is called <u>reciprocity</u>.

Layers

We might have a tower of objects, lower objects less complex than higher ones. It would be nice to be able to say something about a higher object by looking at the just lower object. Given what we know of the elementary particles, might we be able to say something about particles that might appear if the energy were much much higher? Given what we know about integers and primes, might be able to say what would happen if we added in the square root of minus one ( $=i$ )? (Some primes would no longer be prime ($2= (1+i) \times (1-i) = 1 \times 1 - i \times i = 1+1$), and perhaps there would be new primes.)

There is another layer-like phenomena. When we study chemistry, we never have to worry about elementary particle theory, since the characteristic energies are much greater that the energies involved in chemical bonds—million electron-volts vs. one-electron volt. More generally, when we are trying to understand some phenomenon, usually there is a characteristic energy of that phenomenon. Phenomena with much greater characteristic energy are in effect hidden in the setup of the phenomenon, and those phenomena with characteristic energies that are much smaller do not make their appearance (we are insensitive to them in general). We are able to effectively pay attention to the phenomena at hand, the others either hidden in the setup or too small to be seen by us.

Fermions

In chemistry, we learn that each of the electrons surrounding an atomic nucleus has a unique name ($n$, $l$, $j$), describing it "path" around the nucleus. That is why the Periodic Table looks like it does. The electrons are arranged in shells, and within a shell each electron has a unique name besides its shell number. We learn that chemical reactions often depend on electrons in the outer or valence shell. Moreover, in a solid such as a metal, again electrons have unique names and they cannot all bunch up around one nucleus (for then their names could not be unique). Rather, they are "pushed out" to higher energies so to speak, much as racers are distinguished by their race-times.

On the other hand, the photons or light particles between the mirrors of a laser may well have the same names, so that the light is "coherent" and the intensity of the emitted light goes as the square of their number (rather than being proportional to their number). Imagine worshippers in a congregation, singing together, all of whom are equally important. Their prayer will be much louder than were we just to add up each of their prayers when they are alone. (They stimulate each other to sing more strongly.) These are "bosons."

Particles such as electrons and protons and neutrinos and quads are fermions. Particles such as photons, W's, Higgs, and gluons are bosons, and they transmit the force between the fermions.

A Concrete Realization of the Dedekind-Weber-Langlands Program

In 1882 Richard Dedekind and Heinrich Weber showed how (1) Bernhard Riemann's geometric and analysis (continuity) account of algebraic functions of one variable might be proven (2) algebraically using as a model what was then known of (3) algebraic numbers. (Here "algebraic" means defined by a polynomial with integer or polynomial coefficients.) In 1940, Andre Weil pressed the analogy further, as a threefold analogy. And in the last fifty years, the (Robert) Langlands' program might be seen to connect the analytic and automorphic, the algebraic, and the arithmetic. The <u>analogy</u> is how (2)'s theory is made to look like (3)'s, so that (2) provides another theory paralleling (1), but now algebraic rather than "pictorial" or drawing from physics.

In the last eighty years, physicists have been computing the the properties of a two-dimensional lattice of magnetically interacting spins ("The Ising model"), as a classical system, employing a variety of methods. Their goal is to obtain the partition function, PF= $\Sigma$ exp $-\beta E_i$, where $\beta$ is an inverse temperature and $E_I$ is the energy of the lattice for a particular configuration of spins, and the sum is over all such configurations. The thermodynamic free energy, from which the various physical properties of interest might be readily computed, is
$(-1/\beta)$ x ln PF.  The <u>analogy</u> here is between three methods of calculating the same object, coming to the same conclusion, with moments of recognition of an aspect of one calculation mirrored in another.

It would appear that the physicists' computations instantiate what might be called the *Dedekind-Weber-Langlands* program (*DWL*), in effect, an analogy of analogies, what is called a syzygy. One method of computing the partition function of the lattice focussed on the thermal symmetries, such as PF($k$) ~ PF($1/k$), an automorphy or retention of form, where $k$ is a temperature-like variable. A second method was essentially algebraic, in the end finding the right particles or excitations of the lattice (patterned rows of spins), while along the way noting that the $k$-symmetry points to algebraic commutation rules. And a third method might be called arithmetic, finding systematic ways of adding up the contribution of the individual spins for each configuration, in effect a topology of paths.

Keep in mind that the analogy in the mathematician's case is a very different sort of analogy than in the physicist's, for one is an an analogy of theories, and the other is an analogy of methods of calculation (although the theories behind each method might be said to be what is analogized).

Fortunately(!), all methods get the same result. But as in DWL, each method provided insights rather more difficult to infer through the other methods: automorphy, the right particles and a group representation mirroring the automorphy, and an analogy with sum-of-histories formulation of quantum mechanics. More generally, elliptic curves and elliptic functions exhibit the same analogy—having analytic/geometric, algebraic, and arithmetic aspects, intimately connected since they are about the same object, the elliptic curve.

Mathematicians working in the DWL tradition might be encouraged were they to recognize the Ising model solutions as a model for their work—as an analogy to their analogy. My purpose in this brief notice is to bring the syzygy to the fore, while not at all claiming that the physicists' solutions will tell the mathematicians what do. As for the physicists, the DWL program gives coherence to their diverse methods. It's not by chance that there are the three methods, rather they would seem to be more deeply connected (not only in the Ising model, but more generally).

Decisions

Decisions are sometimes transformative, and their value is difficult to anticipate—they are <u>big</u>. There is, as well, <u>uncertainty</u>, where we do not know what we do not know, yet we must act. Acting has consequences that are informative about what to do next, we are never at rest, and sacrifice is the occasion for invention and for revelation of our situation's otherwise hidden aspects.

If we make analytic descriptions of our situation, we will reveal what has been unsaid and hidden. The trick is not to depend on a single description; rather, to have a multiplicity of such from various perspectives or methods, so that we understand our situation as a whole, an *identity in a manifold presentation of profiles*. We will run into blockages, which in their totality reveal further structural features of our situation.

We might think of our situation as a balance of forces—what holds it together, what dissipates it. We might call the most concentrating force *gravity*, only to be stopped by what is intrinsically incompatible ("fermions" cannot be squeezed together too much). There are facticities that cannot be denied. The world having been so compressed—think of a massive hot star—there will be an explosion, a supernova, the ejecta providing surprises. (In this case, heavier elements.)

We find, again, that there are many descriptions, in effect a set of analogies, and there are as well analogies to those sets (what are called syzygies, analogies of analogies). For example, we find that the prime numbers can be thought of in several ways, each way showing different features of those numbers as a whole. We might enhance those numbers (say adding in the square-root of -1) and find that as a whole that enhanced system can be made to be analogous to the original primes (albeit in that enhanced system some of the original primes now prove to be composite). What is remarkable is that we can understand features of the enhanced system from features of the original system of primes.

Moreover, just as kinship systems in a society, prescribe and proscribe marriages and relationships, and that every allowed matching will occur and our labels on people are unique for each person—so is the case for interactions of elementary particles. (There are some qualifications of "unique," but they do not upset my story.)

Just as we can package the primes into a single function, the zeta function, we package the molecules in a room into a function, the partition function, that then leads to the thermodynamic relationship of pressure, volume, temperature, and … Those packaging functions have emergent properties, usually much simpler than keeping track of all of their components. One way of putting this is that if we package the sounds of a drum, we may discern its geometric shape. There are several ways we might build each package, and each of those different modes of packaging reveal features of the package that are hard to discern in the other modes.

There is surprise, multiplicity, and identity, so there is coherence, but there is as well resistance and blockage. If you choose the right individuals, you will find lovely packages, often indicating levels of compositeness of the components of that level. What is remarkable, is that we may discern an order among the objects not at all apparent in their individuality (what is called "reciprocity"), in effect a legacy of the package.